\documentclass[sigplan, 10pt, screen]{acmart}
\usepackage{algorithm}
\usepackage{algpseudocode}

\acmConference[unpublished]{manuscript under preparation}{ 2023}{somewhere}
\setcopyright{none}

\begin{document}

\title{DMC4ML: Data Movement Complexity for Machine Learning}

\author{Chen Ding}
\author{Christopher Kanan}
\author{Dylan McKellips}
\author{Toranosuke Ozawa}
\author{Arian Shahmirza}
\author{Wesley Smith}

\pagestyle{plain}

\begin{abstract}

The greatest demand for today's computing is machine learning. This paper analyzes three machine learning algorithms: transformers, spatial convolution, and FFT.  

The analysis is novel in three aspects.  First, it measures the cost of memory access on an abstract memory hierarchy, instead of traditional time or space complexity.  Second, the analysis is asymptotic and identifies the primary sources of the memory cost.  Finally, the result is symbolic, which can be used to select algorithmic parameters such as the group size in grouped query attention for any dimension size and number of heads and the batch size for batched convolution for any image size and kernel size.

\end{abstract}



\maketitle

\section{Introduction}
Inference in large deep learning models is incredibly expensive and resource-intensive, e.g., ChatGPT is estimated to cost \$700,000 per day to run~\cite{chatGPTExpensive}. Designing methods that are more resource efficient is highly desirable, but using quantitative metrics, e.g., running time and memory, alone can only happen after training the models, which can cost millions of dollars. OpenAI reportedly tried to create a more efficient version of ChatGPT, but they dropped the project because efficiency projections failed to meet expectations~\cite{arrakis}. In this paper we propose new symbolic analysis methods for characterizing the performance of deep learning models without needing to train them first.

\emph{Symbolic analysis} quantifies the cost across machine and input parameters. It complements quantitative testing by helping programmers analyze bottlenecks and write code that is less machine dependent. In contrast to quantitative metrics, symbolic analysis is machine agnostic. Time complexity (TC) is the primary form of symbolic algorithmic analysis; however, it fails to capture the benefits of many optimizations. Consider quick sort. Loop reordering at the program level and memory allocation at the library level do not affect TC because of uniform memory access cost assumption. It is thus oblivious of the memory hierarchy which, especially in data heavy workloads, is critical to performance and energy efficiency. \emph{Data movement complexity (DMC)}, a recently proposed alternative to time complexity (TC), overcomes this weakness by measuring the cost of memory accesses in a mathematically defined geometric memory hierarchy~\citep{DingS:MEMSYS21,Smith+:ICS22}. We demonstrate how DMC can predict benefits of some of the optimizations that are currently being used to make deep learning models more efficient to run and train.

DMC is ordinal in that it is totally ordered.  Past methods of symbolic analysis includes I/O complexity~\citep{HongK:STOC81} and the analysis of cache-oblivious algorithms~\citep{Frigo+:FOCS99} and communication-avoiding algorithms~\citep{You+:THPC21}.  They quantify the cost by the amount of communication for a given local memory size.  It is not ordinal when considering the whole memory hierarchy, because their comparison depends on the cache size.

Unlike time complexity, DMC discerns rather than ignores constant factors.  For example, a factor of two is immaterial in time complexity, but for data movement a factor of two is like the difference in human movement between traveling one-way and round trip.



Using DMC, we analyze machine learning algorithms and quantify their design and optimization.  Till now, most algorithmic analysis is limited to time and space complexity.  The chief novelty of this paper is the asymptotic analysis of a set of algorithms by their memory cost.  



The main contributions of this paper are as follows:
\begin{itemize}
    \item DMC analysis of transformers
    \item DMC analysis of spatial convolution and several of its variants: im2col, Winograd Fast convolution, and batched convolution
    \item DMC analysis of FFT and comparison with spatial convolution
    \item The effect of memory layout, including spatial locality
\end{itemize}

DMC is a theory with its limitations which we discuss in Section~\ref{sec:background}.  
DMC4ML is valuable in at least two ways.  First, the analysis may help in practice to evaluate algorithmic parameters, e.g., the size of batching, based on the abstract memory cost.  Second, optimizing for DMC may improve system design.  For example, a recent study by DeepMind \cite{Fawzi+:Nature22} has reduced the time complexity of matrix multiplication, and another does neural architecture search to improve convolution, with one dimension of optimization being FLOPS \cite{TanL19}. A future study may target the asymptotic or numeric reduction of DMC or use it in tandem.  Hence, DMC could be used by machine learning to optimize machine learning.

    

\section{Background}
\label{sec:background}

\begin{figure}[h!]
    \centering
    \includegraphics[width=6cm]{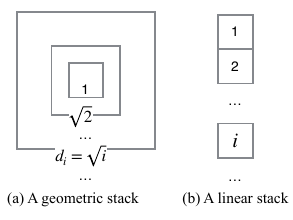}
    \caption{The geometric stack (a) and the linear stack (b).  The data movement distance $d_i$ in the geometric stack is the square root of the stack distance $i$ in the linear stack.}
    \label{fig:geostack}
\end{figure}

Data Movement Complexity (DMC) is a recent theory to quantify the cost of data movement\cite{DingS:MEMSYS21,Smith+:ICS22}.  It models  memory as an abstract cache hierarchy which we call the \emph{geometric stack} shown in Figure~\ref{fig:geostack}(a).  The stack consists of a series of levels numbered $1, 2, \dots$.  Each level $i$ stores one more unit of memory than the previous level, starting with the first level storing at the first unit memory.  

\citet{Mattson+:IBM70} gave the first definition of stack memory, which is a linear stack showing in Figure~\ref{fig:geostack}(b), and defined the \emph{stack distance} for each memory access.  DMC uses a geometric stack and calculates the cost of the memory access by the square root of the stack distance:

\begin{definition}[Data Movement Distance (DMD)]
\label{def:gsd}
For an access sequence $a$ and any stack algorithm, let $l_i$ be the stack distance for the $i$th access $a_i$. Its Data Movement Distance (DMD) $d_i$ is 
\begin{align*}
    d_i = \sqrt{l_i}
\end{align*}
\end{definition}

\begin{definition}[Data Movement Complexity (DMC)]
    The data movement complexity (DMC) of an algorithm is the total DMD of all its memory accesses.
\end{definition}
DMC and the total DMD are synonyms.  We tend to use the former when it is symbolic, i.e., a formula, and the latter when it is numeric.  For DMC, we use the asymptotic equivalence notation $\sim()$, which is identical to big-O notation except it retains primary factor coefficients.

A processor can be considered to reside at level 0.  Its memory starts at the stack level 1.  A smaller stack distance means better locality.  
In \citet{Mattson+:IBM70}, the stack may be managed by any algorithm that has the inclusion property in that the content of a smaller cache is a subset of that of a larger cache.  In this paper, we consider the \emph{LRU stack distance}.  LRU stands for the least recently used replacement policy.  It is the basis of all practical cache designs.




LRU stack distance is also called the \emph{reuse distance}, which is the amount of data accessed since the previous access to the same datum~\citep{Zhong+:TOPLAS09}.\footnote{The reuse distance, or synonymously, the LRU stack distance, is measured asymptotically faster using a tree instead of a stack~\citep{Zhong+:TOPLAS09}.}  In the following analysis, the data movement distance is the square root of the reuse distance. For example, in $abbbca$, the reuse distance of the second $a$ is 3.  Its DMD is $\sqrt{3}$.  

Using DMC, \citet{Smith+:ICS22} analyzed the effect of tiling in matrix multiplication.  While tiling does not change the time complexity, it reduces DMC asymptotically.  


\smallskip
Being a theory, DMC makes a number of assumptions.  The abstract cache hierarchy is an approximation of a real cache hierarchy, just as the abstract cost is an approximation of the real cost.  The theory does not make claims about accuracy with respect to observed numerical performance differences. For example, an algorithm with half the data movement cost compared to another, does not necessarily run twice as fast. 

DMC has no model of parallelism.  It assumes that processing is centrally located and supported by a memory hierarchy.  It does not distinguish whether CPUs or GPUs carry out the processing.  It does not consider distributed caches.  The geometric stack assumes two-dimensional data layout.  The analysis in this paper assumes fully associative LRU caches.

\section{Data Movement Complexity Analyses}

We first derive the DMC of a transformer, spatial convolution and FFT and then discuss these results. 

\subsection{Transformer}
\paragraph{Background}We provide intermediate steps and a complete overview of DMC for a forward and a backward pass through a decoder-only transformer. As DMC is not yet adapted for parallel analysis, we assume a sequential pass through multi-head attention modules. We assume a dropout of .1, as in the original Attention is All You Need paper \cite{attention}, and we follow their presented model for our calculation. The basic algorithm is given below: 
\begin{algorithm}
    \begin{algorithmic}
\Statex $x \gets \text{embed}(input)$ 
\Statex $x \gets x$  + pos$\_$encode $(x)$  \newline
\For{Decoder Layers}
      \State x $\gets$ x + mmha(x) \Comment{Masked Multi-Head Attention}
      \State x $\gets$ norm(x)
      \State x $\gets$  x + mha(x) \Comment{Multi-Head Attention}
      \State x $\gets$ norm(x)
      \State x $\gets$ x + fnn(x) \Comment{Fully Connected Feed Forward\\ \;\;\;\;\;\;\;\;\;\;\;\;\;\;\;\;\;\;\;\;\;\;\;\;\;\;\;\;\;\;\;\;\;\;\;\;\;\;Neural Network}
      \State x $\gets$ norm x
\EndFor
\State x $\gets$ linear(x)
\State x $\gets$ softmax(x)
    \end{algorithmic}
\end{algorithm}\\
The crux of this calculation is DMC for multiplying non square matrices, which closely follows the calculation given by \cite{wesog}. Completing this calculation yields 
\begin{lemma}[DMC of multiplying non-square matrices]
Given  A $\in \text{Mat}_{m \times n}(\mathbf{R})$, B $\in \text{Mat}_{n \times l}(\mathbf{R})$, we have 
    \begin{equation*}
        D_{A \times B}(m,n,l) \sim (m(nl)^{3/2})
    \end{equation*}
\end{lemma}
\paragraph{Multi-Head Attention}
We follow the algorithm for MHA proposed in Attention is All You Need\cite{attention}. We let Q,K,V be the query, key, and value matrices shared by all heads, all of dimension $l \times d$. Then, let $W_Q^i, W_K^i, W_V^i$ be the projection matrices for the $i$th head. Let the query and key projections have dimension $d \times k$, and the value projection  dimension $d \times v$. If there are $h$ heads, we take $k = v = d/h$. Our final weight matrix, $W_0$,  has dimension $hv \times d.$ \begin{algorithm}
    
    \begin{algorithmic}
    \For{Heads}
        \State head $\gets$ softmax$(\frac{Q*W_Q^i * (K * W_K^i)^T }{\sqrt{d_k}} + \text{mask}) * (V*W_V^i)$
    \EndFor
     \State concat(heads) * $W_O$
    \end{algorithmic}
\end{algorithm}We consider one head, and note that the difference between the first head and the following is asymptotically insignificant. 
If we assume $l,h \ll d$, as is the case for most use cases,
 we can further symplify and find the total asympotatic DMC to be roughly 
 \begin{lemma}[DMC For an Attention Head]
      \begin{align*}
    D_{head}(l,v,d,h) 
     &= \sim\bigg( {ld^{3}}(\frac{1}{\sqrt{h}}) + hld^{5/2} + hl^{5/2} + hl^{5/2}d^{3/2}  + hl\bigg)\\
     &= \sim(ld^{5/2}h + \frac{ld^3}{\sqrt{h}})
  \end{align*}
 \end{lemma}

  Note that whether or not the mask is included is asymptotically insignificant. Finally, we consider the DMC contributed from the multiplication of the concatenated head outputs and the weight matrix. This is an $l \times hv$ matrix multiplied by an $hv \times d$ matrix, yielding the result below with substitution $v = \frac{d}{h}$.
  \begin{lemma}[DMC for Multi Head Attention]
      \begin{equation*}
      D_{MHA}(l,v,d,h) \sim(l(hvd)^{3/2}) = (ld^{3})
  \end{equation*}
  \end{lemma}
 Note that, interestingly, this is independent of the number of heads.

\paragraph{Forward Pass}We present, a table of asymptotic equivalences for each stage of the transformer. Important aspects to note are that the DMC of a layer of the transformer decoder model is asymptotically equivalent for the first layer, last layer, and all layers between. We define $d$ to be the dimension of our model, $l$ to be the maximum length of an input token, and $f$ to be the dimension of the Feed Forward Neural Network.\\
\begin{table}[h!]
    \centering
    \begin{tabular}{|c|c|}
    \hline
         Embedding & $(dl)^{3/2}$  \\
         \hline
        Positional Encoding &  $(dl)^{3/2}$ \\
        \hline
        Masked Multi-Head Attention & $ld^3$ \\
        \hline
        Multi Head-Head Attention & $ld^3$\\
        \hline
        Feed Forward Neural Network & $\sqrt{4df}$\\
        \hline
        Linear & $3ld^{3/2}$\\
        \hline
        Softmax &$ 2l^{5/2}$\\
        \hline
    \end{tabular}
    \caption{DMC Transformer Stages}
    \label{tab:DMC Transformer Stages}
\end{table}
This shows that the largest asymptotic term is the multihead attention. Then, if we have $n$ layers, we conclude \begin{theorem}[DMC For Forward pass of a Transformer]
    \begin{equation*}
        D_{forward pass} =\sim (n(2ld^3))
    \end{equation*}
\end{theorem}
\paragraph{Grouped Query Attention (GQA)}
Recently, transformers have been implemented using grouped query attention\cite{gqa}, which shares key and value vectors between groups of attention heads. If we let there be $p$ groups, each of size $q$, we have $pq=h$. The original MHA implementation had $q=1, p=h$, and Multi-query attention has $p=1, q=h$. This causes a performance speed up, despite the cost of compute being the same. This performance speedup can be explained by DMC. We introduce a proposed cost of cold misses, expanded upon in section \ref{sec:confirmation}. \begin{lemma}[Cost for cold miss initialization of data object of size m]
\begin{equation*}
    D_{cold miss} = (m)^{1.5}
\end{equation*}

\end{lemma} 

A cold miss cost is incurred from each time we allocate memory for new matrices in the attention group, that is the query, key, and score matrices. These are size $dk, dk, dd$ respectively. Between each element in the key and value projection matrices there are both of these matrices (size $dk = \frac{d^2}{h}$), the base key and value matrices (size $kd$), the score matrix (size $d^2$) and the value matrix (size $dv = \frac{d^2}{h}$). Square rooting the sum of these yields the reuse between matrices, each of which are size $dv$.  Note that each group reuses a matrix $q-1$ times, as the final iteration has no reuse. We present below how MHA changes with the number of groups, omitting the cost of matrix multiplication as this remains constant and instead focusing on cold misses and $W_Q,W_K$ matrix reuses. 

\begin{lemma}[DMC for Grouped Query Attention] There are $p$ groups of size $q$, where $W_Q,W_K$ are of size $d \times v$, and the score matrix is of size $d \times d$]
\begin{align*}
    D_{\text{GQA}}& =\sim p(d^2 + \frac{2d^2}{h})^{1.5} +  2p(q-1)\frac{d^2}{h} \sqrt{l^2+\frac{4d^2}{h}} \\
   & =\sim pd^{3} 
\end{align*}
\end{lemma}
We observe that the since $p(q-1) \sim h$, DMC grows almost linearly with the number of groups for a fixed $d$. This is because the cold miss cost outweighs the cost of reusing elements within groups, as within groups elements of the key and query matrices are relatively close. \\We investigate how this evaluation can inform model design for transformers. We fix the cost to be $10^5$ DMD, and set head size 8, 32, 64, 128. The given curves advise how to adjust model dimension against the size of groups in MHA to satisfy a cost constraint.

\begin{figure}[h]
    \centering
    \includegraphics[width=8.2cm]{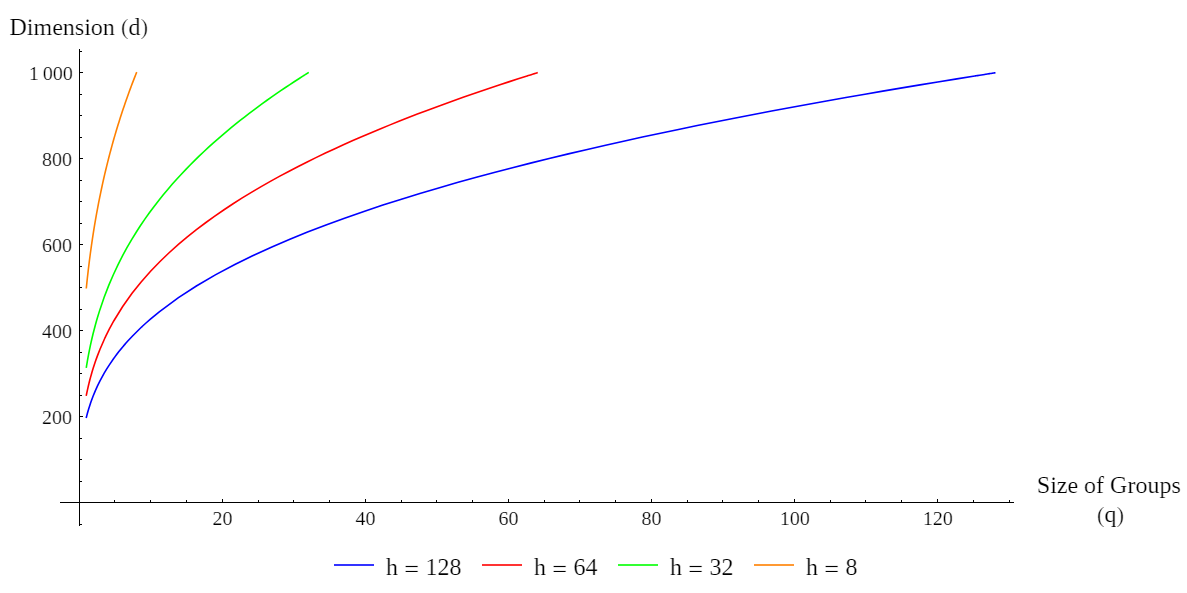}
    \caption{Benefits of GQA measured as the increase in dimension $d$ over the group size $q$ for a fixed cost of $10^5$ DMD. Each curve shows a different number of heads $h$.}
    \label{fig:qroupedreuses}
\end{figure}

\subsection{Spatial Convolution}

A convolution operation takes as input an $I_{n,m}$ image with $n$ rows and $m$ columns and a square kernel $K_{k,k}$, which has $k$ rows and columns.

\paragraph{\textit{Algorithm setup}} Our first analysis is on naive spatial convolution, as its pervasive use makes its performance vital to the efficiency of machine learning applications. Our input setup and algorithm are as follows:
\begin{algorithm}
\caption{Convolution(I[n][n], K[k][k])}\label{alg:cap}
\begin{algorithmic}
\For{$i = 0 \ldots (n - k + 1)$}
\For{$j = 0 \ldots (n - k + 1)$}
\State sum = 0
\For{$y = 0 \ldots k$}
\For{$x = 0 \ldots k$}
\State sum += K[y][x]$\ast$I[i+y][j+x]
\EndFor
\EndFor
\State R[i][j] = sum
\EndFor
\EndFor
\State return R
\end{algorithmic}
\end{algorithm}
\begin{equation*}
I_{n, n} = 
\begin{pmatrix}
I_{1,1} & I_{1,2} & \cdots & I_{1,n} \\
I_{2,1} & I_{2,2} & \cdots & I_{2,n} \\
\vdots  & \vdots  & \ddots & \vdots  \\
I_{n,1} & I_{n,2} & \cdots & I_{n,n} 
\end{pmatrix}
\end{equation*}
\begin{equation*}
K_{k, k} = 
\begin{pmatrix}
K_{1,1} & \cdots & K_{1,k} \\
\vdots  & \ddots & \vdots  \\
K_{k,1} & \cdots & K_{k,k} 
\end{pmatrix}
\end{equation*}

From here on out, we refer to the 2D input as an `image'. For our initial examination, we assume the kernel overlap with the matrix is bounded by the borders of the image, that is, the image is already padded with zeros if it is needed. We will derive the result for square images, but for readability and space constraints will present the result for non-square images without derivation.

\paragraph{\textit{Kernel and Row-wise Reuses}} We break down reuses into different categories. As a general pattern, we think of naive spatial convolution as a sequence of windows, each a single iteration within the nested $i$ and $j$ for loops.  First we examine reuses of $k$[][], kernel elements, and row-wise reuses of the image. 
\begin{align*}
W_{1, 1} = 
\begin{pmatrix}
K_{1,1} * I_{1,1} & \textcolor{blue}{K_{1,2}}  *  \textcolor{red}{I_{1,2}}  & \cdots & K_{1,k} * I_{1,k}\\
\vdots  & \vdots & \ddots & \vdots  \\
K_{k,1} * I_{k,1} & K_{k,2} * I_{k,2} & \cdots & K_{k,k} * I_{k,k}
\end{pmatrix}\\
W_{1, 2} = 
\begin{pmatrix}
K_{1,1} * \textcolor{red}{I_{1,2}} & \textcolor{blue}{K_{1,2}} *  I_{1,3}  & \cdots & K_{1,k} * I_{1,k+1}\\
\vdots  & \vdots & \ddots & \vdots  \\
K_{k,1} * I_{k,2} & K_{k,2} * I_{k,3} & \cdots & K_{k,k} * I_{k,k+1}
\end{pmatrix} 
\end{align*}

The matrices above demonstrate the reuses between the first and second sliding windows. This can be extended to consider any two consecutive sliding windows on the same row. Within any reuse interval of an element of the kernel, $K_{1, 2}$ in this case, every element of the kernel will be accessed. In addition, every element of the first window (except for $I_{1, 1}$) will be accessed. As for how many times this reuse pattern occurs, it is simply the number of kernel elements ($k^2$) multiplied by the number of windows ($(n-m + 1)^2 - 1$). This leads to the following: 
\begin{lemma}[Convolution: Kernel Reuses] 
\label{lemma:kernel}
Let $f(n, k)$ be the data movement complexity of kernel reuses in convolution with image size $n$ and kernel size $k$. Then 
\begin{align*}
 f(n,k) = k^2((n-k+1)^2-1)\sqrt{2k^2-1}
 \nonumber
\end{align*}  
\end{lemma}
A similar pattern can be seen for the reuse interval of the image element. The biggest difference is that the number of row-wise reuses varies for different elements as well as some consecutive windows.
This is because after $(n-k+1)$ windows, the next window will be on a new row. Consequently, elements in the corners and on the rightmost column will not have row wise reuses. This leads to the following:
\begin{lemma}[Convolution: Row-wise Reuses] 
Let $g(n, k)$ be the DMC of row-wise reuses in convolution with image size $n$ and kernel size $k$. Then 
\begin{align*}
    g(n,k) = k(k-1)(n - 3k + 2)(n-k)\sqrt{2k^2 - 1} 
 \nonumber
\end{align*}  
\end{lemma}
Another way to view this equation is that most windows have $k(k-1)$ row wise reuses, and that there are $(n - 2k + 2)(n-k)$ windows where this applies. Note that this equation also does not count the windows $k-1$ rows away from the bottom and top row, which do have reuses, but simply occur less often and are asymptotically insignificant. 

\paragraph{\textit{Column-wise Reuses}} The final set of reuses we examine are reuses of image elements in different window rows (different $i$ in the pseudocode), which we refer to here and in our full derivation as column-wise reuses. One way of characterizing column-wise reuses is that they occur for image elements which are paired with the rightmost edge of a kernel at some point during a sliding window row. This means that there are $(k-1)$ column-wise reuses per window, $(n-k-1)^2 - 1$ windows as before, and (ignoring asymptotically insignificant constants) each reuse contains $(n\cdot k)$ image elements and $k^2$ kernel elements.
\begin{lemma}[Convolution: Column-wise Reuses]
    Let $h(n, k)$ be the DMC of column-wise reuses in convolution with image size $n$ and kernel size $k$. Then
    \begin{align*}
        h(n,k) = (k-1)((n - k - 1)^2 - 1)\sqrt{n\cdot k + k^2}
    \nonumber
    \end{align*}
\end{lemma}
After combining all three DMC terms, and eliminating irrelevant ones, we arrive at
\begin{theorem}[Convolution Data Movement Complexity]
\label{thm:wh-conv}
Let $D_{conv}(n, k)$ represent the DMC of spatial convolution with a kernel of size $k$ and a square image of size $n$, and let $D'_{conv}(h, w, k)$ be the same but with an $h \times w$ image. Then
\begin{align*}
D_{conv}(n,k) &= \sim(2\sqrt{2}k^3n^2 + k^{1.5}n^{2.5})\\
D'_{conv}(h,w,k) &= \sim(2\sqrt{2}k^3hw + k^{1.5}hw^{1.5})
\end{align*}
\end{theorem}
\noindent where $k$ is the kernel size, $n, w, h$ the square and non-square image width and height.

\paragraph{Width and Height Asymmetry}
An image has a height-to-width ratio which is one for square images, more than one for portraits, and less than one for landscapes.
If the image size is far greater than the kernel size, i.e., $w,h \gg k$, the second term dominates.  In the second term, the cost grows faster with the width $w$ than with the height $h$.  Compared to portrait images of a height-to-width ratio $\frac{h}{w}=m$, the data movement is a factor of $\sqrt[4]{m}$ and $\sqrt{m}$ larger for square and landscape images with the same number of pixels.  The difference is 19\% and 41\% respectively for $m=2$.




\subsection{Convolution Variants}
\label{sec:conv-variants}
\paragraph{\textit{Batching}} 
Consider image processing where a third dimension is the number of color channels used for the image and there is a kernel for each channel.  The question for batching
is this: is it better to process each channel individually, or ``batch" multiple channels at once?  We consider an extension of the naive convolution scheme where there are $c$ channels and there is an option to perform the convolution in multiple passes, each time performing convolution on $x$ channels. This means the convolution will be performed in $\lceil\frac{c}{x}\rceil$ batches. This yields the following:

\begin{theorem}[Batched Convolution]
\label{thm:batch-conv}
Let $D_{batch}(n, k, c, x)$ represent the DMC of spatial convolution with a kernel of size $k$, a square image of size $n$, $c$ data channels, and batch size $x$. Then
\begin{align*}
D_{batch}(n, k, c, x) &= \\ \sim(c\sqrt{x}(2\sqrt{2}k^3n^2 + k^{1.5}n^{2.5}) + \sqrt{x}(\frac{c}{x} - 1)n^3)
\end{align*}
\end{theorem}

\noindent The high level reasoning is as follows.  By batching $x$ convolutions at a time, each reuse is being inflated by a factor of $\sqrt{x}$. This accounts for the first two terms, which come from $c\sqrt{x}\cdot D_{conv}(n,k)$. The last term is the cost of adding results from all channels to the result matrix.  There is just one access and no data reuse when $c=1$; otherwise, when there is no batching, each result element has $c-1$ reuses at reuse distance $n^2$, and when the batching factor is $x$, the number of batches is $\frac{c}{x}$ (assuming $c$ is divisible by $x$ for simplicity), each result element is reused $\frac{c}{x}-1$ times at reuse distance $n^2x$.



Without batching, convolution is performed on a single channel and faster, while the result matrix is updated once per channel.  With batching, convolution uses more memory, but the result update once per batch.  This trade off is shown in $D_{batch}(n, k, c, x)$, whose first term increases and second term decreases with batching.  The total shows the combined effect.

\begin{figure}[htp]
    \centering
    \includegraphics[width=8.4cm]{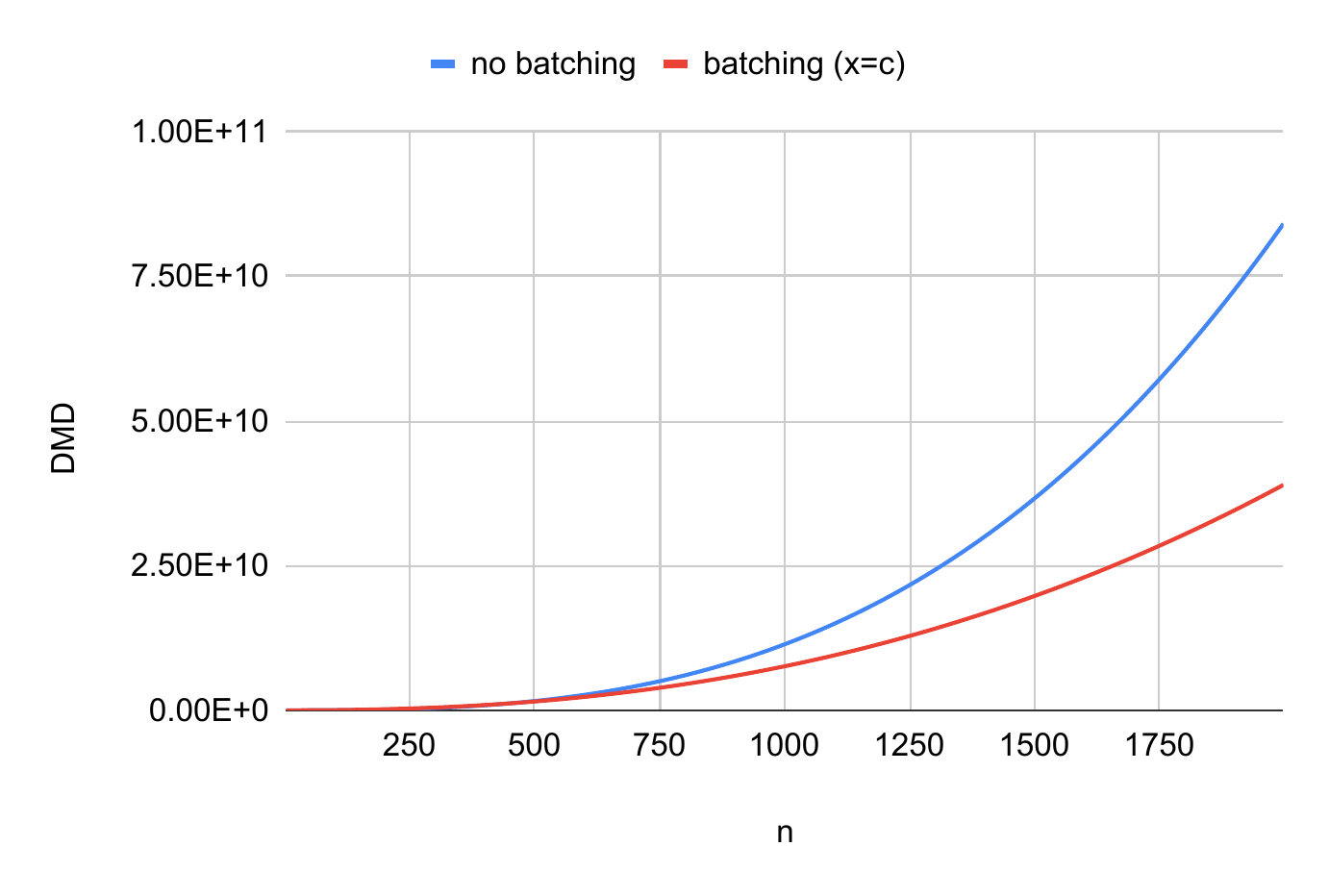}
    \caption{Batching effect shown by the total DMD across image sizes $n$, with kernel size $k=3$ and number of channels $c=10$.}
    \label{fig:batching-n}
\end{figure}

Figure \ref{fig:batching-n} shows the effect of batching as a function of image size $n$.  The kernel size is 3, and there are 10 channels. Batching uses $x=c=10$.  For data sizes smaller than 448, batching hurts DMD, but with larger data sizes, batching reduces the DMD. When data size increases from 448 to 2000 (364\% data increase), the savings from batching changes from a 0\% to a 47\% decrease in DMD.

\begin{figure}[htp]
    \centering
    \includegraphics[width=8.4cm]{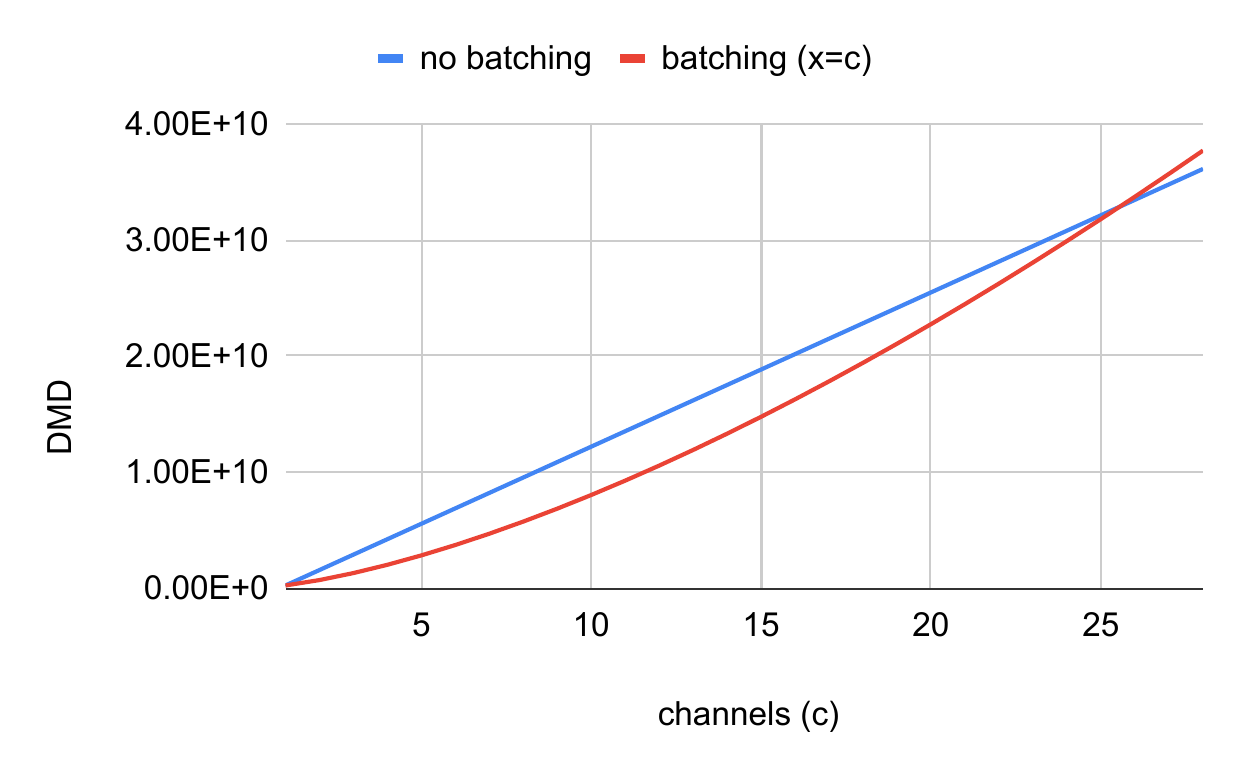}
    \caption{Batching effect shown by the total DMD over different degrees of batching, with image size $n=1024$ and kernel size $k=3$.}
    \label{fig:batching-c}
\end{figure}

Figure~\ref{fig:batching-c} shows the effect with a different degree of batching.  Here the image size is fixed $n=1024$, with kernel size $k=3$.  The figure shows the total DMD across the number of channels $c$.  All channels are processed in one batch, i.e. $x=c$.  Before $c=25$, batching is beneficial; after it, batching is not.  The equation $D_{batch}(n, k, c, x)$ lets us compute when batching is too much given $n,k,c$.
\paragraph{\textit{im2col}} We also examined im2col, a prepossessing step to vectorize spatial convolution, on DMC. In order to represent spatial convolution as a matrix-vector multiplication, one has to somehow convert each window into a column. We considered a modification of the spatial convolution algorithm to do this conversion. 

\begin{algorithm}
\caption{Im2Col(I[n][n], k)}\label{alg:cap}
\begin{algorithmic}
\For{$i = 0 \ldots (n - k + 1)$}
\For{$j = 0 \ldots (n - k + 1)$}
\State p = 0
\For{$y = 0 \ldots k$}
\For{$x = 0 \ldots k$}
\State R[j+p][i] = I[i+y][j+x]
\State p += 1
\EndFor
\EndFor
\EndFor
\EndFor
\State return R
\end{algorithmic}
\end{algorithm}

From here, doing a matrix vector multiplication with the kernel as a vector has the same reuse pattern as the kernel reuses described in the derivation of spatial convolution (Lemma \ref{lemma:kernel}). In terms of accesses to the image, the reuse pattern in the im2col process is extremely similar. The only difference is that elements of the kernel are no longer accessed (which was asymptotically irrelevant in the original analysis anyways). A key difference, however, is there are now reuses of elements of $R$, the matrix form of the convolution. The full derivation is in the tech report, and yields the following:

\begin{theorem}[im2col Data Movement Complexity]
\label{thm:im2col}
Let $D_{im2col}(n, k)$ represent the DMC of convolution performed via the im2col procedure. Then
\begin{align*}
D_{im2col}(n,k) &= \sim(2\sqrt{2}k^3n^2 + k^{1.5}n^{2.5} + kn^{3})
\end{align*}
\end{theorem}

It is important to note that the last term, which represents the additional cost of im2col over spatial convolution, is incurred because we are not considering parallel computation. In fact, the actual pre-processing step of im2col, represented by the first two terms, incurs no additional. And although this result does not fully explain the performance benefit of im2col, it helps expose the tradeoffs made. In exchange for better parallelism, there is a cost associated with creating a larger matrix and having those elements be stored in cache over the duration of the process.

The last term asymptotically dominates when $n >> k$. However, it represents only a factor of $\frac{\sqrt{n}}{\sqrt{k}}$ increase: for practical sizes (e.g. $n=1024, k=7$) this is around 12x. While substantial, this is an increase that can be recouped by the data-level parallelism allowed by the technique: modern Intel processors, for example, allow SIMD instructions on 16 32-bit data at once. Moreover since parallel computation would entail more cores and caches for each, essentially reduce every reuse distance would be reduced in addition to the parallel speedup. Thus the actual amount of recovery possible is even greater.

\paragraph{\textit{Winograd Fast Convolution}}
Winograd Fast Convolution \citep{winogradFast} is an optimized form of convolution that improves performance via an alternative way to compute matrix-vector multiplication for specific problem sizes. It is similar to Strassen's algorithm, but instead of reducing the recursive branching factor (which reduces the overall time complexity of the algorithm) it reduces the number of expensive multiplication operations needed, making it a purely compute optimization, not a data movement based optimization.

\subsection{FFT}
In this section we derive the DMC of the Fast Fourier Transform (FFT). Due to space constraints, we do not provide the full derivation, but instead the high-level structure, with some intermediate results presented without a comprehensive proof.
\paragraph{\textit{Algorithm Setup}} For our analysis of the FFT, we examine the standard recursive algorithm as follows:

\begin{algorithm}
	\caption{FFT(n, A: [$a_{0}, a_{1}... a_{n-1}$])}\label{alg:cap}
	\begin{algorithmic}
	\If{n==1}
	        \State return A
	\EndIf
	\State $F_{even}$ = FFT$(n/2, [a_{0}, a_{2}... a_{n-2}])$
	\State $F_{odd}$ = FFT$(n/2, [a_{1}, a_{3}... a_{n-1}])$
	\For {$k=0\ldots n/2$}
		\State Y[k] = $F_{even}$[k] + $\omega[\frac {k}{n}]$ * $F_{odd}$[k]
		\State Y[k + $\frac{n}{2}$] = $F_{even}$[k] - $\omega[\frac {k}{n}]$ * $F_{odd}$[k]
	\EndFor
	\State return Y
	\end{algorithmic} 
\end{algorithm}
\noindent where $n$ corresponds to the number of elements in $A$, and is a power of 2. The $\omega$ represents a global dictionary or array which is indexed by the loop variable $k/n$, each element being $e^{2 \pi ik/n}$. For the sake of clarity, when referring to elements of the structure we will simply use the induction variable $k$ in the top level call. For instance, suppose the root call is of size 16, then the roots of unity accessed at each level of the recursion will be as follows:
\begin{align*}
\text{FFT(16)}: \omega[0], \omega[1], \omega[2], \omega[3]... \omega[7] \\
\text{FFT(8)}: \omega[0], \omega[2], \omega[4], \omega[6] \\
\text{FFT(4)}: \omega[0], \omega[4] \\
\text{FFT(2)}: \omega[0] 
\end{align*}
The memory used to store these precomputed values is of no concern, as the analysis is not taking into account spatial locality. 
\vspace{-0.2cm}
\paragraph{\textit{Divide and Conquer Phases}} The computations performed by the FFT create temporary variables between each recursive call. We distinguish between reuses of temporary variables which are created in the ``divide phase" of the algorithm and the ``conquer phase" of the algorithm, the high level pattern for which can be shown below:

\begin{figure}[htp]
    \centering
    \includegraphics[width=8.5cm]{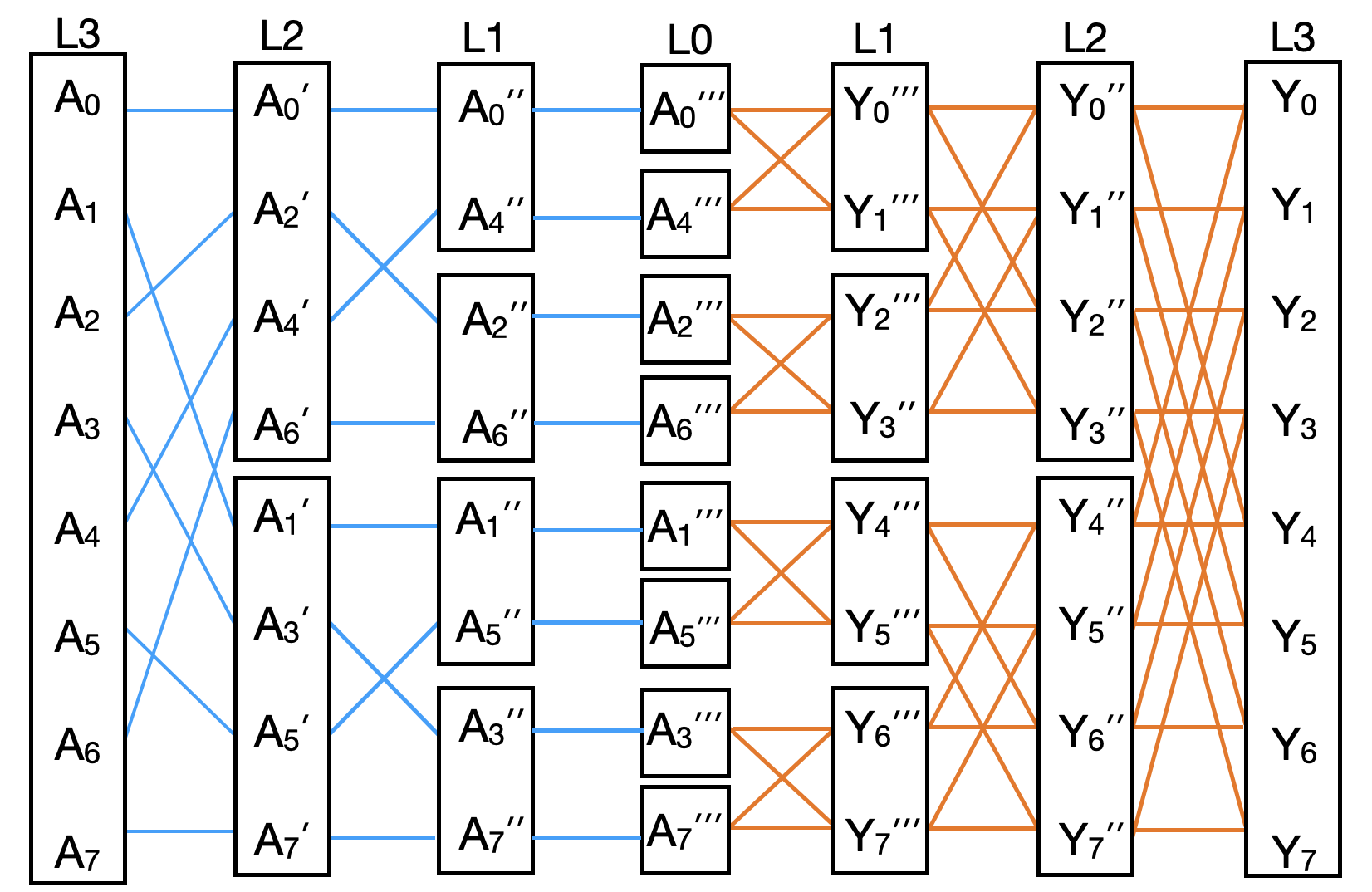}
    \caption{FFT phases. The blue lines denote the ``Divide" phase, and the orange the ``Conquer" phase.}
    \label{fig:Butterfly-diagram}
\end{figure}
The blue represents temporaries during the divide phase of the algorithm and involve memory accesses when elements of $A$ are being copied into subsequent recursive calls. The orange represents the memory accesses and computation associated with the creation of temporaries during the conquer phase of the algorithm, when elements of $Y$ from a lower depth (level) recursive call are being used to compute $Y$ in the current recursive call. 

It is clear from this diagram and pseudocode that every temporary will only be accessed in a call which is at most one level different from where it is created. Thus it is useful to determine the number of elements accessed by a recursive call of arbitrary size $N$, in FFT($N$, $A$). This yields a closed form function:

\begin{lemma}[FFT: Data size by level] Let $F(L)$ be the number of unique elements a recursive call at level L accesses, where L $\in \left[ 0, \log_2(N)\right]$. 
\begin{align*}
F(L) = (2L + 1.5) \cdot 2^L
\end{align*}  
\end{lemma}

This is because the total number of temporaries which persist throughout a call to FFT($N$, $A$) is always $2N$ at any given level (as shown in Figure \ref{fig:Butterfly-diagram}). The $1.5$ term represents the fact there are always N elements at level 0, and $2^{L-1}$ elements of $\omega[]$ to consider. 

The takeaway here is that we can treat a recursive call FFT($N$, $A$) as a black box from the perspective of memory if all we are interested in are reuse distances, as we know the amount of data the function call will access.

\paragraph{\textit{Divide Phase Reuses}} The divide phase of the algorithm can be summarized as constantly splitting A in FFT($N$, $A$) into subsequent calls by even and odd indexed elements. Even indexed elements are copied  immediately as a FFT() call starts, while copying the odd indexed elements has one whole recursive call in between. This means the Divide Phase Reuses of the odd indexed elements are asymptotically greater than those of the the even indexed elements, which leads to the following:

\begin{lemma}[FFT: Divide phase DMC] 
\label{fft:div}
Let $f(n)$ be the DMC incurred by divide phase reuses of the FFT with input size $n$. 
\begin{align*}
f(n) = \sum_{d=2}^{\log_{2}(n)} 2^{\log_{2}(n) - d} \cdot \sum_{a=0}^{2^{d-1}}  \sqrt{F(d-1) + C(a)} 
 \nonumber
\end{align*}  
\end{lemma}

For the sake of space, we use $C(x)$ to refer to elements which are accessed out of recursive calls captured by $F(L)$ (loop structures and array iteration for example) throughout this brief derivation. These elements have no effect on the asymptotic analysis, and can be upper bounded as a small constant multiple of $N$.

\paragraph{\textit{Conquer Phase Reuses}} Here we examine reuses which occur during the conquer phase of the algorithm. In each recursive call, elements of $Y$ are returned to their parent call, assigned to either $F_{odd}$ or $F_{even}$, and used to compute $Y$ values in the parent. Similar to how the Divide reuses of even indexed elements happen immediately, right after $F_{odd}$ is assigned, the loop in the pseudocode starts and their elements from the previous $Y$ call are used almost immediately. On the other hand $F_{even}$ elements have asymptotically larger reuse distances because they have to wait for multiple recursive calls before they are accessed again. 

\begin{lemma}[FFT: Conquer phase DMC]
\label{fft:conq}
Let $g(n)$ be the DMC incurred by conquer phase reuses of the FFT with input size $n$. Then 
\begin{align*}
 g(n) = \sum_{d=2}^{\log_{2}(n)} 2^{\log_{2}(n) - d} \cdot
\sum_{a=0}^{2^{d-1}}  \sqrt{F(d-1) + C(a)} 
 \nonumber
\end{align*}  
\end{lemma}

For both Lemmas \ref{fft:div} and \ref{fft:conq}, the interpretation for the summations are simple. The terms inside the square root are the reuse distances. The first summation is iterating over every level; the factor of $2^{\log_{2}(n) - d}$ is counting the number of recursive calls at level $d$, and the second summation is iterating over each (relevant) element in a singular call.

\begin{figure}[htp]
    \centering
    \includegraphics[width=8.1cm]{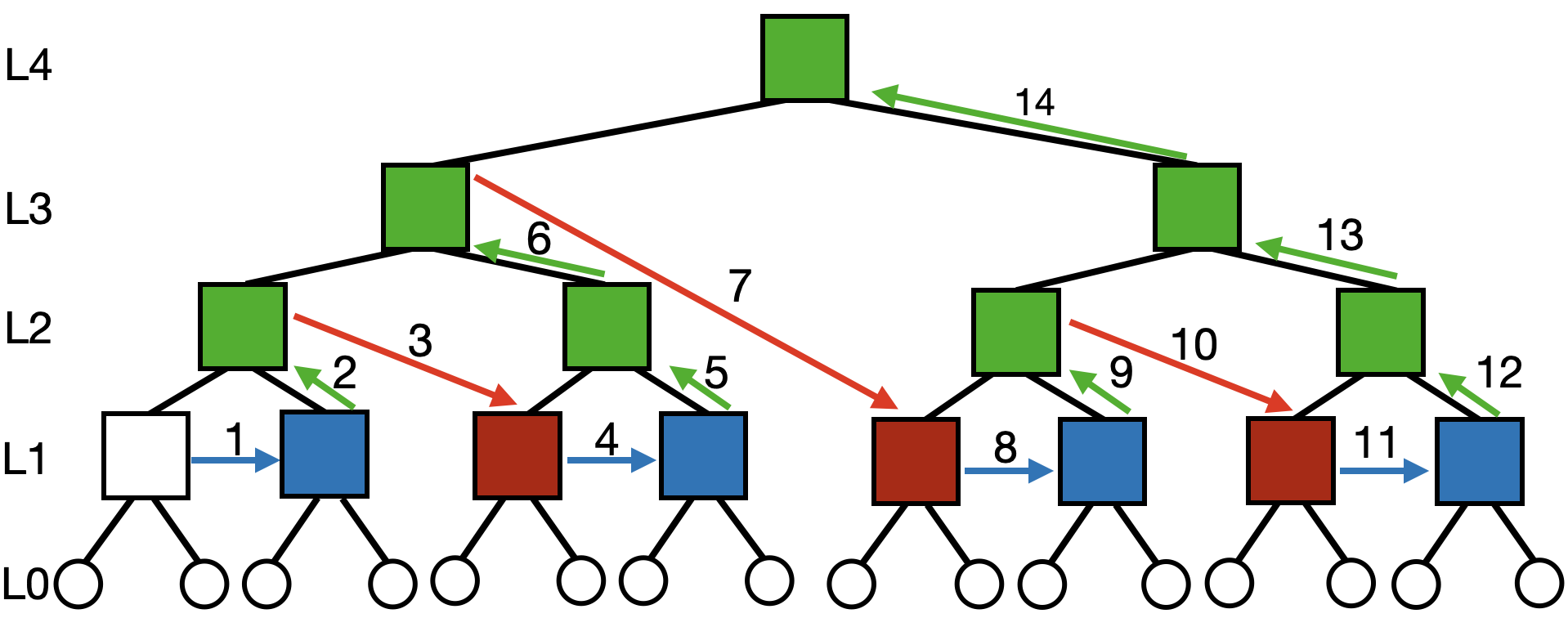}
    \caption{$\omega[0]$ reuses for FFT(16). Distant reuses in red, child-parent reuses in green, and sibling reuses in blue.}
    \label{fig:omega reuses}
\end{figure}

\paragraph{\textit{Roots of Unity Reuses}} The final, and most complex, set of reuses come from the repeated memory accesses to elements of $\omega$, i.e. the roots of unity. Not only do these reuses span multiple levels of recursive calls, they also span across branches of the tree which represents the FFT's execution. We first examine memory accesses of the ``0th" root of unity root, i.e. $\omega[0]$. This element appears in every recursive call, aside from the base case when $N = 1$. Its reuses are shown in Figure~\ref{fig:omega reuses}. 

 The previous two types of reuse were homogeneous and happened at every level of recursion, however that is not the case here. And so it is first useful to derive the frequency distribution of these types of reuses before looking at reuse distances themselves. 

\paragraph{\textit{Distant Reuses}} The most noticeable feature of the tree is that sibling reuses only happen at L1, and so to count the sibling reuses we simply have to count the number of L1 node pairs, or the number of L2 nodes. This is simply $\frac{n}{4}$. 

From there we observe that there is a one to one correspondence between the number of sibling pairs (not reuses) and the number parent reuses. There is also an exact correspondence between the sibling pairs and the number of distant and sibling reuses combined. This gives us the following:

\begin{lemma}[FFT: Distance Reuse Count] Let $c(0, n)$ be the number of distant reuses for $\omega[0]$ for FFT(n, A). 
\begin{align*}
 c(0, n) = \sum_{d=2}^{log_{2}(n)-1} 2^{log_{2}(n)-1-d} = \frac{n}{4} - 1
\end{align*}  
\end{lemma}

At a high level, the summation is counting at every level $d$ the number of sibling pairs there are. Since the sibling pairs at level 1 are simply sibling reuses, we start at level 2. 

This also tells us that there are $\frac{n}{2}-1$ child-parent reuses. However, child-parent reuses, similar to the $F_{odd}$ conquer phase reuses, only access elements within loop structures. There are no recursive calls in between any child-parent reuse and likewise it does not affect the asymptotic result. 

To generalize our reuse analysis of $\omega[0]$ to any element of $\omega$, we first have to determine why elements do not appear in every recursive call. For starters, every odd index element has no reuses and only appears at the root call. 
\begin{align*}
\text{FFT(16)}: \omega[0], \omega[1], \omega[2], \omega[3]... \omega[7] \\
\text{FFT(8)}: \omega[0], \omega[2], \omega[4], \omega[6] \\
\text{FFT(4)}: \omega[0], \omega[4] \\
\text{FFT(2)}: \omega[0] \\
\end{align*}

\begin{figure}[htp]
    \centering
    \includegraphics[width=8.2cm]{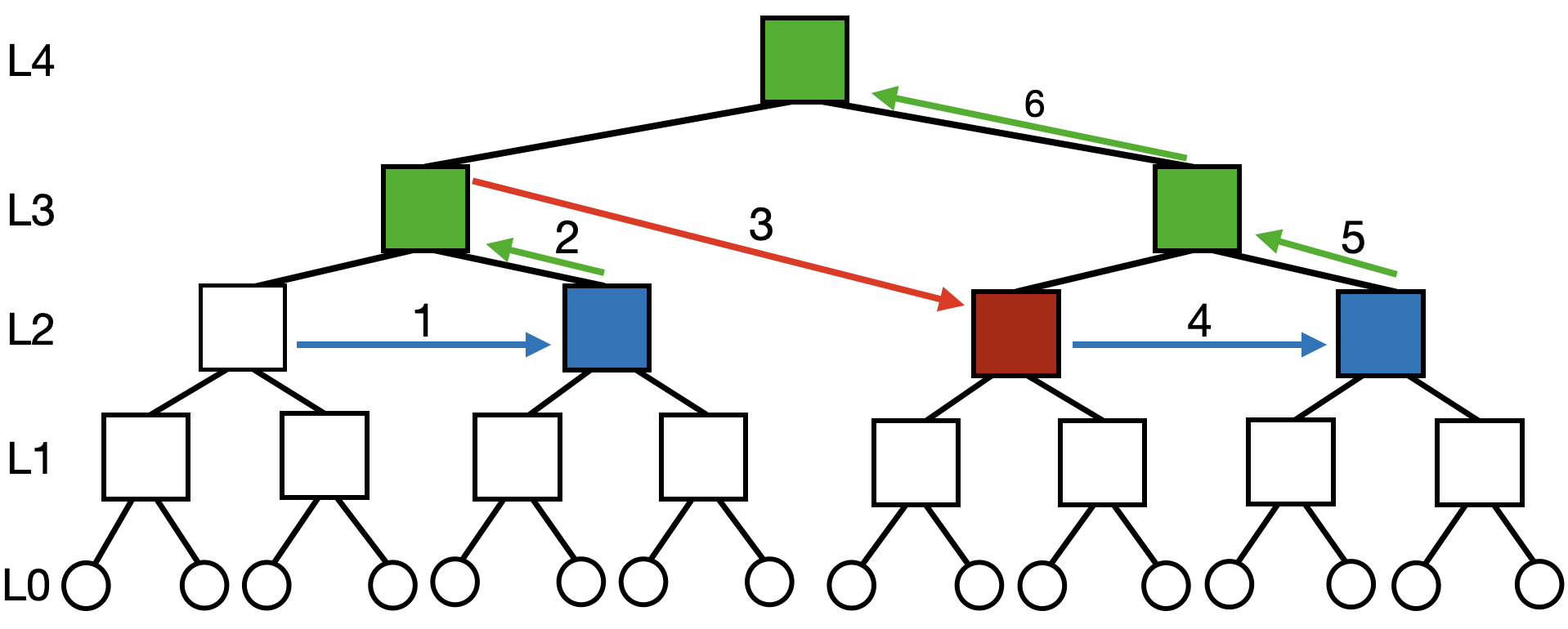}
    \caption{$\omega[4]$ reuses for FFT(16). Distant reuses in red, Parent reuses in green, and Sibling reuses in blue.}
    \label{fig:omega reuses 2}
\end{figure}

From there, every element of $\omega[]$ which lands at an odd $k$ in the loop structure will only have sister reuses.  Figure~\ref{fig:omega reuses 2} shows such reuses for $\omega[4]$.

This means that for any given $\omega[b]$, the depth that it will appear in the tree will correspond to how many 2s there are in the prime factorization of $b$. Combining this information with the previous lemma, we arrive at the following:

\begin{lemma}[FFT: Generalized Distant Reuse Count] Let $c(n, b, X)$ be the number of distant reuses for some non zero b in $\omega[b]$ given FFT($n$, $A$), where X is the lowest level in which b appears. 
\begin{align*}
 c(n, b, X) = \sum_{d=X+1}^{log_{2}(n)-1} 2^{log_{2}(n)-1-d} = \frac{n}{2^{X+1}} - 1
\end{align*}  
\end{lemma}

The actual reuses distances for Sibling and Distant reuses are asymptotically similar to the previous lemmas, which we omit here for brevity.
 


After taking into account every DMC component, we can construct upper and lower bounds. 

\begin{theorem}[FFT: DMC Bounds] Let $D_{FFT}(n)$ be the DMC of the FFT with input size $n$. Then
\begin{align*}
D_{FFT}(n) = f(n) + g(n) + h(n) + k(n) \\
D_{FFT}(n) \geq \sim(6.4n^{1.5} \sqrt{\log n}) \\ 
\leq \sim(6.5 n^{1.5} \sqrt{\log n})
\end{align*}  
\end{theorem}
Our upper and lower bounds are tight to within 0.1 on the leading coefficient, and are asymptotically tight.

\subsection{Discussion}
\subsubsection{\textit{FFT for Convolution}} 
\label{sec:fft4conv}
FFT can be used to compute convolution.
The first step is to extend the FFT result to 2D inputs. Performing a FFT on a 2D input requires performing a FFT on each row of the input, followed by performing it on each column (or vice versa). To perform FFT convolution, one has to take the 2D FFT of a padded kernel and the original image, take the point-wise product between the two results, and then perform the inverse FFT on the the final result. The cost of the inverse FFT is the same as the normal FFT, as there is no difference in memory behavior. This allows us to construct a generous, but asymptotically tight, lower bound as follows:
\begin{align*}
D_{FFTconv}(n) &\geq \sim(3* 2n * D_{FFT}(n))\\
& \geq \sim(38.5n^{2.5} \sqrt{\log n})
\end{align*}
\noindent where $n$ is the dimension of the square image, and also is a power of 2. For comparison, the cost of kernel-based convolution is:
\begin{align*}
D_{conv}(n,k) &= \sim(2\sqrt{2}k^3n^2 + k^{1.5}n^{2.5})
\end{align*}
\noindent where $k$ is the kernel size, $n$ the image size.

If the image size is far greater than the kernel size, i.e., $n \gg k$, spatial convolution is more efficient than FFT by a factor proportional to $\sqrt{\log n}$.

If $k^3 > n$, the cost of FFT convolution is actually asymptotically smaller, as the first term in convolution ($2\sqrt{2}k^3n^2$) dominates. Since common image sizes are up to $500 \times 500$ or higher and common kernel sizes 7 or lower, usually $n > k^3$, and FFT is less memory efficient for convolution.  Numerically, FFT costs 376 million for $n=512$, while the cost of convolution is 51, 159, and 364 million for kernel dimensions $3,5,7$ respectively.  The gap increases as image size increases. 





\subsubsection{Memory Layout and Spatial Locality}
\label{sec:spatial-locality}
\paragraph{\textit{Data Granularity}}
Data movement complexity until now is quantified with abstract ``data size" - the cost of a memory access is a function of the number of distinct data items accessed. In practice, this can be better quantified by considering the size (in bits) of the data items in question. This yields the following:
\begin{definition}[Granular Data Movement Distance]
\label{def:dmd-gran}
For a program $p$ with data accesses $a_i$ to $s$-bit data, let the reuse distance of $a_i$ be $d_i$.  The DMD is
\begin{align*}
    DMD_{s}(p) = \sum_i \sqrt{s \cdot d_i}
\end{align*}
\end{definition}
The $\sqrt{s}$ distributes out of the summation, and we see that any algorithm's DMC when considering data granularity is $\sqrt{s}$ times that without.

This constant factor reduction is significant enough to warrant industry attention and optimization: much attention has been paid in recent years to deep learning using reduced-precision floats for energy and memory efficiency.
\paragraph{\textit{Spatial Locality}}
A related issue is that of spatial locality - modern memory systems do not cache individual data, but instead cache blocks.
For naive convolution with $b$-element cache blocks, the DMC is as follows:
\begin{theorem}[Convolution DMC with Spatial Locality]
\label{thm:conv-spatial}
Let $b$ be the number of data items in one cache block and assume that both the image and the kernel are laid out in row-major order. Then the DMC of convolution, quantified with cache blocks instead of data items, is
\begin{align*}
    \sim(D_{conv}(n, k)) = \frac{1}{\sqrt{b}}(2\sqrt{2}n^2k^3 + n^{2.5}k^{1.5}) 
\end{align*}
\end{theorem}

Note that the improvement (a factor of $\frac{1}{\sqrt{b}}$) suggested by the above analysis only persists because our analysis does not consider data size, which would add a factor of $\sqrt{b}$ and negate this gain. 
The effect of spatial locality is not to reduce DMC, but to prevent block-based memory layout from incurring \textit{additional} data movement by utilizing each data item present in a block that is loaded into cache. This additional cost is bounded as follows:
\begin{theorem}[Block Layout Cost Bound]
    Let $b$ be the number of data items in one cache block and let the DMC (without considering block layout) of algorithm $A$ be $f(A)$. 
    \begin{align*}
        DMC_b(A) < \sqrt{b} \cdot f(A)
    \end{align*}
\end{theorem}
The proof of this theorem is trivial: padding each data item to the size of a block with useless data that is never accessed during execution yields this result.

\paragraph{\textit{Channels Last Optimization}}
A memory-based optimization for convolution is Channels Last. Its main benefit is that it allows convolution algorithms that process multiple channels to benefit from spatial locality. As the number of channels trends upward, convolution on top of a Channels Last memory layout approaches the best possible spatial locality, i.e. every memory access is to a block already in cache. As discussed before, spatial locality does not directly yield an improvement in DMC, but given a memory system that uses cache blocks, taking advantage of spatial locality prevents DMC from increasing, i.e., from having unused data in a cache block which adds data transfer and reduces usable space in cache.

\paragraph{\textit{Cold Misses}}
Some algorithms have data movement costs not captured by DMC as previously described. This is because there is a cost associated with loading data on its first access, i.e., \emph{cold misses}, or memory accesses that cannot result in cache hits~\citep{Hill:Dissertation}. 

A simple way to characterize the cold miss cost for an algorithm is to create a lower bound by noting that, to load $m$ pieces of distant data, each of them must have been at least $m$ units away from the processor before the initial load. This yields a cold miss cost of $m \cdot \sqrt{m}$ for an algorithm that accesses $m$ data.

\subsubsection{Confirmations and A Conjecture}
\label{sec:confirmation}

DMC counts the theoretical cost from memory access at all levels of an ideal cache hierarchy, which is not physical and not amenable to conventional empirical evaluation.  However, its results are consistent with practice in the following cases:
\begin{enumerate}
    \item Batching is always beneficial for spatial convolution (Theorem~\ref{thm:batch-conv}).
    \item The memory cost of im2col pre-processing is not large for practical image and kernel sizes (Theorem~\ref{thm:im2col}).
    \item The Winograd algorithm is beneficial in reducing expensive operations but not memory cost (Section~\ref{sec:conv-variants}).
    \item For practical image and kernel sizes, spatial convolution is more efficient than FFT  (Section~\ref{sec:fft4conv}).
\end{enumerate}

Finally, the convolution cost is asymmetrical between the length and the width of an image (Theorem~\ref{thm:wh-conv}).  The asymmetry is not yet reported in practice as far as we know.

\section{Related Work}
\label{sec:rel}

\citet{HongK:STOC81} pioneered the study of lower bound I/O complexity, measuring memory complexity by the amount of data transfer and deriving this complexity symbolically as a function of the memory size and the problem size.  The same complexity measures were used in the study of cache oblivious algorithms~\citep{Frigo+:FOCS99} and communication-avoiding algorithms~\citep{You+:THPC21}. \citet{olivry+PLDI20} introduces a compiler technique to statically derive I/O complexity bounds. They use asymptotic complexity ($\sim$), much as we do, to consider constant-factor performance differences, while the rest do not. I/O complexity suffers from an issue inherited from miss ratio curves: it is not ordinal, and a comparison depends on the cache size.

Ordinality is useful for algorithm optimization.  The classic example, Strassen for matrix multiplication, targets time complexity which is ordinal in asymptotic terms. The recent approach by AlphaTensor takes the approach of machine learning~\cite{Fawzi+:Nature22}, which may be adapted to optimize for memory.  Such adaptation is easier with ordinality.

\citet{Callahan+:JPDC88} defined the balance between computing and memory costs.  It was later extended to multi-level caches~\citep{DingK:JPDC04} and by the roofline model~\citep{Williams+:PC09} and its extensions such as a model of heterogeneity by \citet{Gysi+:ICS15}.  Balance and roofline models 
require caches of known sizes (and not shared with other programs), while DMC aims to measure the effectiveness of optimization across all cache sizes. 



Memory hierarchies in practice may vary in many ways, which make a unified cost model difficult.  \citet{Valiant:ESA08} defined a bridging model, Multi-BSP, for a multi-core memory hierarchy with a set of parameters including the number of levels and the memory size and three other factors at each level. A simpler model was the uniform memory hierarchy (UMH) by \citet{Alpern+:UMH94} who used a single scaling factor for the capacity and the access cost across all levels.  Both are models of memory, where caching is not considered beyond the point that the memory may be so implemented. Another important early idea quantifying memory access cost is Memory logP \cite{CameronS:IPDPS03}, which extends the LogP model of parallel computation to consider a hierachical memory subsystem.

How to optimize convolution systems has been extensively researched \cite{Iandola+ICIP13, chen2016eyeriss, shah2018runtime, lin2017data}, however much of the focus is on the computation aspect as opposed to the memory aspect. Recently, \citet{Chen+HPCA20} derive the I/O complexity of convolution 
and argue that the energy cost of convolution is memory bound. \citet{Ivanov+:MLSys21} optimized the implementation to minimize the data movement in the transformer algorithm for a given hardware system.

Neural Architecture Search is a widely used technique for automating the design of neural network architectures~\cite{elsken2019neural}. It has been used to identify architectures that excel in a range of hardware settings, especially for embedded systems, where the search was constrained to optimize machine learning metrics (e.g., accuracy) and FLOPS~\cite{howard2019searching,tan2019mnasnet}. However, neural architecture search is expensive, with many methods requiring thousands of GPU-days. While this is improving with algorithmic advances~\cite{wu2019fbnet}, DMC is a complementary approach that could be used to inform the search strategy in these algorithms.

\vspace{-0.2cm}
\section{Summary}
Using DMC, this paper is the first analysis of transformer, convolution, and FFT algorithms for their memory cost.  
The analysis is asymptotic and symbolic, which can be used to select algorithmic parameters such as the group size in grouped query attention for any dimension size and number of heads and the batch size for batched convolution for any image size and kernel size.
It helps to 
understand interactions between algorithms and memory systems to aid algorithm design and optimization, and possibly as an objective function for ML-driven architecture search and algorithm optimization.



\bibliographystyle{ACM-Reference-Format}
\bibliography{all, ml, extra}

\appendix

\end{document}